\documentclass[%
 reprint,
superscriptaddress,
groupedaddress,
showpacs,
showkeys,
 amsmath,amssymb,
 aps,
]{revtex4-1}

\usepackage[utf8]{inputenc} 
\usepackage{graphicx}
\usepackage{subfigure}
\usepackage{dcolumn}
\usepackage{bm}
\usepackage{hyperref}
\usepackage{epstopdf} 

\newcommand{\vett}[1]{\bm {#1}}
\newcommand{\cc}{^{\ast}}

\begin{document}

\preprint{APS/123-QED}

\title{Effective dissipation and nonlocality induced by nonparaxiality}

\author{Nicola Bulso}
\email{nicola.bulso@gmail.com}
\affiliation{Dipartimento di Fisica, Università di Roma, La Sapienza, Piazzale A. Moro 2, I-00185, Roma, Italy}

\author{Claudio Conti}
\email{claudio.conti@roma1.infn.it}
\homepage{http://www.complexlight.org}
\affiliation{Dipartimento di Fisica, Università di Roma, La Sapienza, Piazzale A. Moro 2, I-00185, Roma, Italy}


\begin{abstract}
We investigate beam diffraction and spatial modulation instability of coherent light beams propa\-gating in the non-paraxial regime in a nonlinear Kerr medium. 
We study the instability of plane wave solutions in terms of the degree of non-paraxiality, beyond the regime
of validity of the nonlinear Schroedinger equation. We also numerically analyze the way non-paraxial terms breaks the integrability 
and affect the periodical evolution of higher order soliton solutions. We discuss non-locality and effective dissipation introduced by 
the non-linear coupling with evanescent waves.
\end{abstract}


                             
\maketitle


\section{\label{sec:Introduction}Introduction}
The nonlinear Schroedinger equation (NLS) describes propagation of a light beam in a nonlinear Kerr medium within the paraxial approximation  \cite{TrilloBook, KivsharBook}, and models a large variety of phenomena, like solitons and shock waves, which are routinely experimentally
observed \cite{Encyclopedia, Gentilini2013}. 
The NLS is an integrable equation \cite{Zakharov} that also finds many applications in hydrodynamics, plasma physics, 
and Bose-Einstein condensation, and this promotes the interdisciplinar exchange of information between different fields of nonlinear physics. 
In particular, modulation instability (MI) was first discovered in hydrodynamics and named Benjamin-Feir instability \cite{Feir, StoriaMI}, 
 a mechanism responsible for the breaking of periodic surface waves on an inviscid fluid layer in deep water\cite{Feir}.
MI concerns the instabi\-lity of the plane wave solution of the NLS equation with respect to a sinusoidal perturbation \cite{Agrawal, Newell}, 
and in the framework of high power laser beam propagation corresponds to the spatial modulation of the transverse beam profile 
and the resulting generation of filaments \cite{Campillo, Mamaev}. 
In these respects, MI is often indicated as a precursor to the formation of bright solitons. 
In the time domain, i.e. in the case of a laser beam in an optical fiber, 
MI induces a temporal intensity modulation and generations of trains of light pulses \cite{Agrawal, Newell}, 
and is also relevant during supercontinuum generation \cite{Bandelow2005}.

When considering the spatial nonlinear dynamics of spatial beams,
the limit of application of the NLS, and of the corresponding analysis of MI,
 is represented by the paraxial approximation: the beam propagation must occur in a narrow cone of wavevectors 
around the direction of propagation; this corresponds to a regime in which 
the transverse dimension of the beam is much greater than the beam wavelength\cite{Agrawal, Newell}.
Various investigations show that the paraxial regime may be broken during nonlinear evolution, \cite{Conti11, ContiPRL04, Conti05PRL, Alberucci:11} and this may be extremely relevant for applications as ultra-high resolution microscopy \cite{Barsi2013}.

Many efforts have been made in the direction of develo\-ping more accurate models to go beyond the paraxial approximation \cite{Lax, Feit, Fibich, DiPorto1997, Chamorro:98, Fuente, Ciattoni:00April, Ciattoni:00May, Ciattoni:02, Matuszewski, Moloney2004, Ciattoni:06, Wang}.
In this paper we follow the approach of Kolesik and Moloney, and employ the Unidirectional Pulse Propagation Equation (UPPE)\cite{Moloney2004, Moloney2012} in the monochromatic regime and for a one-dimensional spatial beam.
 The UPPE generalizes the standard NLS and provides a good description of beams whose size could be smaller, or of the same order, of the wavelength and maintains a scalar form. 
Investigating MI by the UPPE can represent a test for such a model for non-paraxial light beam propagating 
in a nonlinear Kerr medium. 

In the following, we introduce the UPPE and study the propagation of one-dimensional spatial light beams in the monochromatic regime. 
We define the degree of non-paraxiality of the beam, and show that in a linear medium 
this equation provides a more accurate description of the diffraction of a light beam with respect to the paraxial evolution equation (i.e., the linear Schroedinger equation), because it includes all higher order diffraction terms and takes into account of the effective dissipation of the energy carried out in the direction of propagation due to the existence of evanescent waves.
We first study the diffraction of non-paraxial Gaussian beams searching for corrections to the well known results from 
linear optics, and show that diffraction can be analytically treated at any order of non-paraxiality.
We then analyze MI: we first consider a modified NLS in which we include all the higher order diffraction terms and calculate the rate of exponential growth for each Fourier component of an initial small perturbation of the transverse beam profile; then, in a later section, we consider the entire UPPE model retaining the corrections to the nonlinear terms. We demonstrate the importance of these corrections and show the way non paraxiality affects the growth of the perturbation. Finally, we numerically investigated the way this higher order terms lead to the breaking of the periodicity of higher order soliton solutions. 
\section{\label{sec:The model}Unidirectional pulse propagation equation}
The UPPE \cite{Moloney2004} is a scalar envelope equation for the transverse components of the electric field and 
furnishes a connection between the nonlinear Maxwell equations and the simple NLS, which is valid in the limit of paraxial propagation. The UPPE is derived from Maxwell equations 
under the valitidy of (i) the scalar, and (ii) the unidirectional approximation. 

The scalar approximation (i) neglects the longitudinal component of the electric field 
and of the polarization vector. Even in a simple isotropic nonlinear medium, nonlinearity couples components of the eletric field, and neglecting the longitudinal components of the electric field and of the polarization vector constitutes an approximation, which is known to be effective in many practical application. 
The validity of the scalar approximation is lost for strongly non-paraxial beams,
 when is necessary to include the longitudinal component of the electric field in the model and take into account 
the full vectorial electric field\cite{DiPorto1997, Fuente, Ciattoni:00April, Matuszewski, Moloney2004}. 
For beams that propagate in a paraxial or slightly non-paraxial regime, the UPPE provides a more accurate description of the beam evolution than the NLS equation and maintains a scalar form. 
 
The unidirectional approximation (ii) neglects the backward propagating component of wave to obtain a closed equation for the forward-propagating part. The error in doing such an approximation is small when the complex amplitude of the forward field is slowly varying with respect to wavelength, a condition valid for slightly non-paraxial propagation\cite{Moloney2004,Moloney2012}.

For monochromatic, linearly polarized light beams, the electric field is written as
\begin{equation}
\vett E(\vett r,t) = \frac{\hat{\vett x}}{2}\left(E(\vett x_{\perp},z) e^{-i\omega_{0} t} + c. c.\right)\text{,}
\end{equation}
and the UPPE in an instantaneous Kerr medium is
\begin{equation}\label{eqn:UPPE_E}
\left(i\partial_{z} + \sqrt{\beta_{0}^{2}-{|\vett k_{\perp}|}^2}\right) {\cal E} = 
- \frac{n_{0} n_{2}k_{0}^{2}}{\sqrt{\beta_{0}^{2}-{|\vett k_{\perp}|}^2}}{\cal P}[E]. 
\end{equation}
In (\ref{eqn:UPPE_E}), ${\cal E}(\vett k_{\perp},z)$ and ${\cal P}[E](\vett k_{\perp},z)$ are, respectively, the Fourier transform with respect to the transverse spatial coordinate, $\vett x_{\perp}$, of the electric field $E(\vett x_{\perp},z)$ and of $|E(\vett x_{\perp},z)|^{2}E(\vett x_{\perp},z)$, being $n_{0}$ the linear refractive index, $n_{2}$ the Kerr coefficient, $k_{0}$ the va\-cuum propagation constant and $\beta_{0} = n_{0}k_{0}$. By factoring out the slowly varying part of the complex amplitude, 
${\cal A} =\sqrt{\frac{\epsilon_{0}n_{0}c}{2}\sigma^2} {\cal E} e^{-i\beta_{0}z}$, with ${\cal A}$ measured in $W^{1/2}$ and $n_{2}$ in $m^2W^{-1}$, we obtain 
\begin{equation}\label{eqn:UPPE_A}
\left(i\partial_{z} + \sqrt{\beta_{0}^{2}-{|\vett k_{\perp}|}^2}-\beta_{0}\right) {\cal A} = -\eta \frac{\beta_{0}}{\sqrt{\beta_{0}^{2}-{|\vett k_{\perp}|}^2}}{\cal P}[A],
\end{equation}
with $\eta = n_{2}k_{0}/\sigma^{2}$ and $\sigma$ is the characteristic transverse size of the beam.
We normalize the transverse variable $\vett x_{\perp}$ to $\sigma$, the propagation variable $z$ to the diffraction length ($L_{d} = \beta_{0}\sigma^2$) and the field amplitude ${\cal A}$ to the peak power $P_{0}$ so that $\hat{\psi} = {\cal A}/\sqrt{P_{0}}$. We define a nonlinear length, $L_{nl} = 1/\eta P_{0}$, and a nonlinear coefficient $\gamma = L_{d}/L_{nl}$ measuring the relative strength of the linear and nonlinear terms. 
Correspondingly, UPPE is rewritten as
\begin{equation}\label{eqn:UPPE_Psi}
\left[i\partial_{z} + \frac{1}{\epsilon}\left(\sqrt{1-\epsilon{|\vett k_{\perp}|}^2}-1\right)\right] \hat{\psi} +\frac{\gamma}{\sqrt{1-\epsilon{|\vett k_{\perp}|}^2}}{\cal P}[\psi] = 0,
\end{equation}
where ${\cal P}[\psi]$ is the Fourier transform of $|\psi|^2\psi$ and we introduce the non-paraxial parameter $\epsilon$ which is defined as $\epsilon = 1/\beta_{0}L_{d}$.
For $\epsilon\rightarrow 0$ , Eq.(\ref{eqn:UPPE_Psi}) reduces to the standard NLS equation. 
The non-paraxial parameter $\epsilon$ determines the strength of the corrections to the paraxial propagation. 
Because of the scalar and unidirectional approximations, highly non-paraxial beams are not described by Eq.(\ref{eqn:UPPE_Psi}); nevertheless, this simple scalar model allows to theoretically investigate light beams with a transverse size slightly smaller or, of the same order of the wavelength, namely $\epsilon \alt 0.1$.
Non-paraxiality natural introduces nonlocality in the model as outlined by Eq.(\ref{eqn:UPPE_Psi}).

\section{\label{sec:diffraction}Nonparaxial diffraction}
As a first application of the UPPE, 
we study the diffraction of a beam propagating in a linear medium ($\gamma = 0$). As mentioned above, the UPPE provides a more accurate description than the linear Schroedinger equation for two reasons: it involves all the higher order diffraction terms and includes the evanescent part of the Fourier spectrum. 
In the linear case Eq.(\ref{eqn:UPPE_Psi}) is
\begin{equation}\label{eqn:UPPE_lineare}
\left[i\partial_{z} + \frac{1}{\epsilon}\left(\sqrt{1-\epsilon{|\vett k_{\perp}|}^2}-1\right)\right] \hat{\psi} = 0.
\end{equation}
In what follows, for the sake of simplicity, we consider the 1+1 dimensional case; the extension to the 2+1 case is straightforward. The solution of Eq.(\ref{eqn:UPPE_lineare}) is given by
\begin{equation}\label{eqn:solution}
\psi (x,z) = \int\frac{dk}{2\pi}\hat{\psi}_{0}(k)e^{ikx + iR(k)z},
\end{equation}
with $R(k) = \frac{1}{\epsilon}\left(\sqrt{1  - \epsilon k^2} - 1\right)$ and $\hat{\psi}_{0}(k) = \hat{\psi}(k,0)$. In order to study the spreading of a light beam, we use the root mean square deviation of $|\psi (x,z)|^2$:
\begin{equation}\label{eqn:rms}
\sigma^2(z) = \frac{1}{\cal N}\int dx\, x^2\, |\psi (x,z)|^2,
\end{equation}
with the normalization constant ${\cal N} = \int dx\, |\psi (x,z)|^2 = \int\frac{dk}{2\pi}\, |\hat{\psi}(k,z)|^2 = \int\frac{dk}{2\pi}\, |\hat{\psi}_{0} (k)|^2$. 
Using Eq.(\ref{eqn:solution}) in (\ref{eqn:rms}):
\begin{eqnarray}\label{eqn:sigma}
\displaystyle\sigma^2(z) &=& \frac{1}{\cal N}\int \frac{dk}{2\pi}\left[ R'^{\,2}| \hat{\psi}_{0}|^2 z^2 -i(R''|\hat{\psi}_{0}|^2 + 2R'\hat{\psi}'_{0}\hat{\psi}^{\ast}_{0})z +\right.\nonumber\\
\displaystyle                 &   & \left. -\hat{\psi}''_{0}\hat{\psi}^{\ast}_{0}\right]e^{-2 R_{i} z},
\end{eqnarray}
where the quotes stands for derivatives with respect to $k$.
$R_{i}(k)$ is the imaginary part of $R(k)$ and is different from zero only for $k > 1/\sqrt{\epsilon}$, 
which gives the evanescent spe\-ctrum. 
Typically $\epsilon \ll 1$, so that the evanescent Fourier components corresponds to $k \gg 1$; 
as the wave has a spectral range of some units in $k$ because of the normalization, the values of $\hat{\psi_{0}}(k)$ for $k \gg 1$ are negligibly small and, correspondingly, the exponential term in equation (\ref{eqn:sigma}) is negligible. 
It turns out that the beam broadens with a quadratic power law along the propagation coordinate
\begin{equation}
\sigma^2(z)  = A z^2 + B z + C,
\end{equation}
with
\begin{eqnarray}
\displaystyle A &=& \frac{1}{\cal N}\int \frac{dk}{2\pi} R'^{\,2}|\hat{\psi}_{0}|^2,\label{eqn:A}\\
\displaystyle B &=& -i\frac{1}{\cal N}\int \frac{dk}{2\pi} (R''|\hat{\psi}_{0}|^2 + 2R'\hat{\psi}'_{0}\hat{\psi}^{\ast}_{0}), \label{eqn:B}\\
\displaystyle C &=& -\frac{1}{\cal N}\int \frac{dk}{2\pi} \hat{\psi}''_{0}\hat{\psi}^{\ast}_{0}.\label{eqn:C}
\end{eqnarray}
We note that the functional dependence is the same as that derived in paraxial approximation. The difference is enclosed in the values of the coefficients $A$ and $B$, which depend on the non-paraxial parameter $\epsilon$. The meaning of the various coefficients is the following. Given that  $\sigma^2(0) = \frac{1}{\cal N}\int dx\, x^2\, |\psi (x,0)|^2 = -\frac{1}{\cal N}\int \frac{dk}{2\pi} \hat{\psi}''_{0}\hat{\psi}^{\ast}_{0}$, $C$ represents the size of the beam at the starting point $z = 0$. The coefficient $B$ is related to the initial chirp of the beam, it is different from zero only if $\hat{\psi_{0}}(k)$ has a nonzero imaginary part. The last coefficient, $A$, specifies the spatial scale for the beam broadening and can be written as 
\begin{equation}\label{eqn:A2}
A = \frac{1}{\cal N}\int \frac{dk}{2\pi}\, \frac{k^2}{1-\epsilon k^2}\,\hat{\psi}^{2}_{0}(k). 
\end{equation}

In order to find an explicit mathematical expression for the evolution of the r.m.s. of the beam, we consider an unchirped Gaussian beam as initial condition: $\psi(x,0) = e^{- \frac{x^2}{2}}/\sqrt{2\pi}$.
We then have $B = 0$ and the integral in (\ref{eqn:A2}) can be evaluated by an asymptotic power series expansion in $\epsilon$. 
We obtain the following expressions
\begin{equation}\label{eqn:otticagaussiana}
\sigma^2(z) = \frac{1}{2}+Az^2,
\end{equation}
with
\begin{equation}\label{eqn:serieasintotica}
A \simeq \sum_{n=0} \frac{(2n+1)!!}{2^{n+1}}\epsilon^{n} = \frac{1}{2} + \frac{3}{4}\epsilon + \frac{15}{8}\epsilon^2 + O(\epsilon^3).
\end{equation}
If the power expansion in (\ref{eqn:serieasintotica}) is stopped to $O(1)$, the well-known results
from linear Gaussian optics is retrieved. At higher orders, it turns out that non-paraxial effects cause a faster broadening of the beam.

\begin{figure}[t]
\centering
\subfigure{\includegraphics[width = \columnwidth]{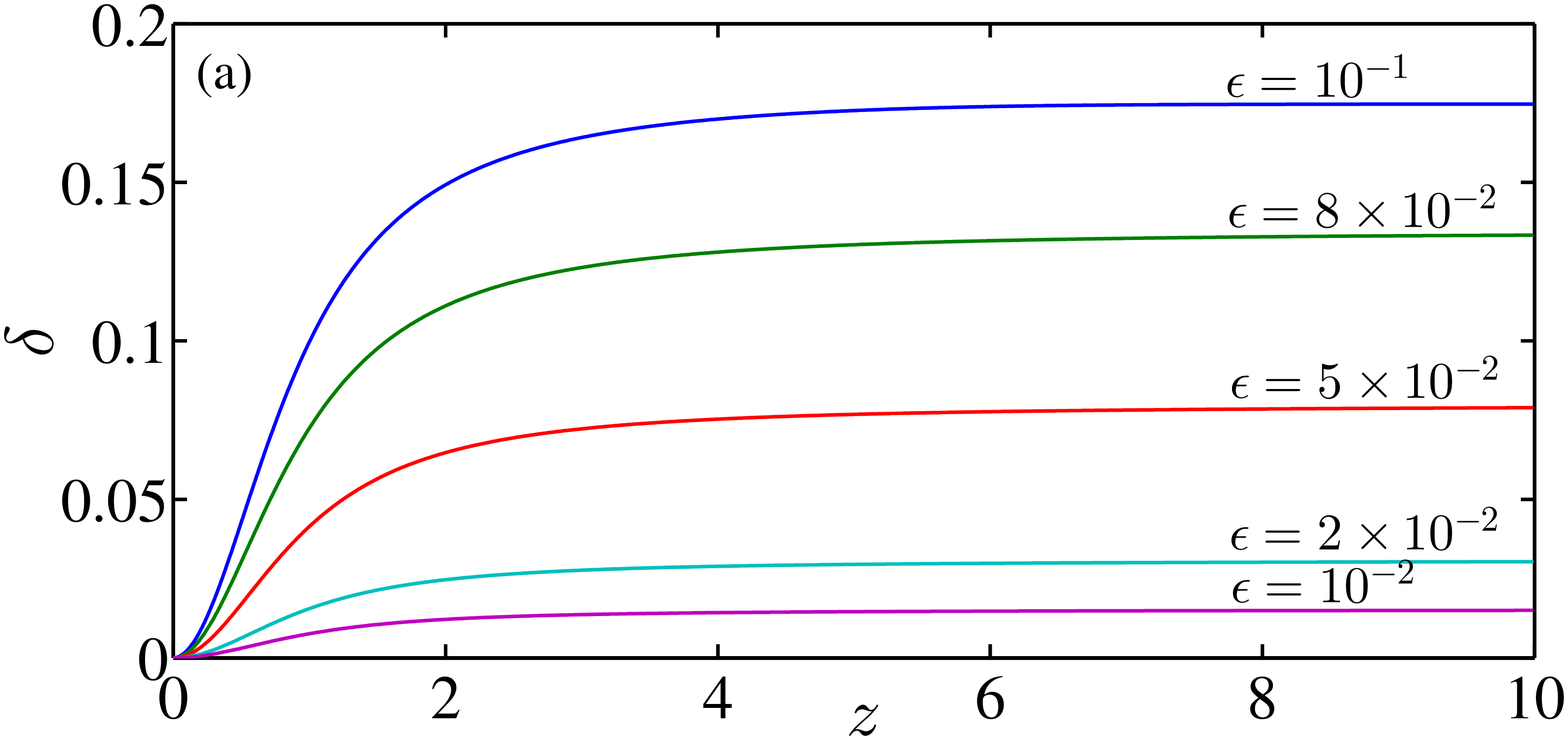}}
\subfigure{\includegraphics[width = \columnwidth]{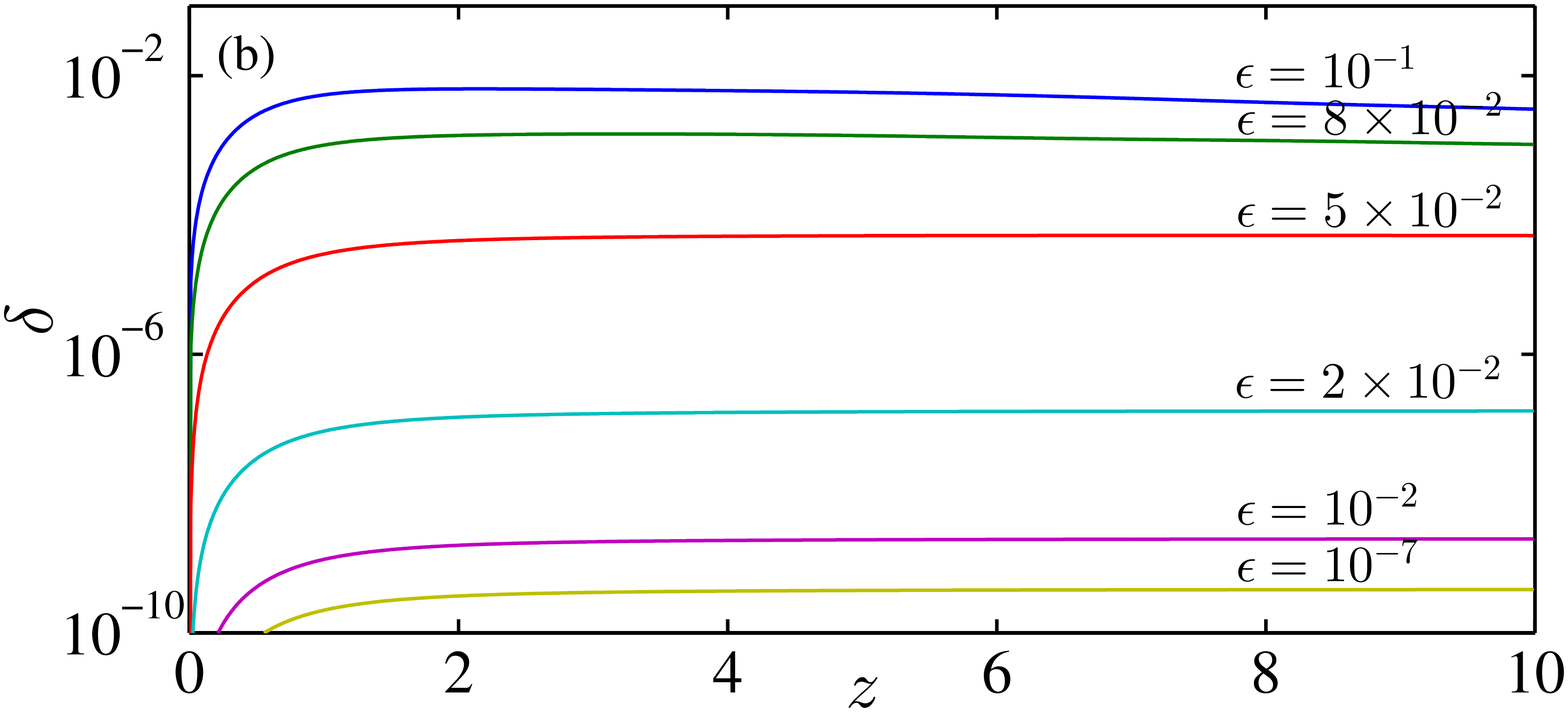}}
\caption{(Color online) (a) Comparison of the numerical si\-mulations of the linear UPPE (\ref{eqn:UPPE_lineare}) with the paraxial theory, given by Eq.(\ref{eqn:otticagaussiana}) with $A$ approximated by the first term in the asymptotic expansion (\ref{eqn:serieasintotica}). The figure shows that the theory is correct up to an uncertainty $\delta$ of $17\%$ for $\epsilon = 0.1$. (b) as in (a) going beyond the paraxial approximation, i.e., including higher order corrections in $\epsilon$ (we use the first six term in the series expansion); in this case the error made is always lower than $1\%$.}\label{fig:confronto diffrazione}
\end{figure}

We compare these theoretical results with numerical simulations of Eq.(\ref{eqn:UPPE_lineare}). 
We found a full agreement between theory and simulations, as shown in Fig.\ref{fig:confronto diffrazione} where $\delta$ represents the relative difference between the theoretical values of the r.m.s. $\sigma_{t}$ 
and those numerically calculated $\sigma_{n}$, namely $\delta = |\sigma_{t} - \sigma_{n}|/\sigma_{t}$.
$\delta$ is of the order of $1\%$ in the worst case for $\epsilon = 0.1$ 
and for a normalized propagating distance $z = 10$; the corresponding value in the paraxial approximation
is about $17\%$. We also note that the discrepancy between paraxial and non-paraxial mo\-dels becomes relevant,
i.e. greater than $1\%$, for $\epsilon \agt 0.01$ corresponding to $\lambda\alt \sigma/2$.
\section{\label{sec:MI}Modulation instability}
We analyze non-paraxial MI by first considering a modified NLS including only the corrections to the li\-near term, i.e. higher order diffraction terms. In a later section, we also retain the corrections to the nonlinear terms, which introduce small nonlocal interactions. As above, to simplify the notation and for the comparison with the numerical simulations we limit to one-transverse spatial dimension, the extension of our theoretical ana\-lysis to two transverse directions is straightforward.
\subsection{\label{sec:Corrections to the linear term}Corrections to the linear term}
We consider the following modified NLS 
\begin{equation}\label{eqn:modifiedNLS}
\left[i\partial_{z} + \frac{1}{\epsilon}\left(\sqrt{1  + \epsilon \partial^{2}_{x}} - 1\right)\right] {\psi} + \gamma |\psi|^2\psi = 0.
\end{equation}
In passing from the Fourier space to the real space we expand the square root in a Taylor series, this is justified by the smallness of the $\epsilon$ parameter, which makes negligible the weight of the evanescent Fourier components.
MI arises from the instability of the exact plane wave solution with respect to an oscillatory perturbation; 
the plane wave solution is given by
\begin{equation}\label{eqn:solondapiana}
\psi (x,z)= e^{i\gamma z}.
\end{equation}
The perturbation to (\ref{eqn:solondapiana}) with wave-number $k$ is
\begin{eqnarray}
\displaystyle\psi (x,z) &=& \left[1+a(x,z)\right] e^{i\gamma z}\label{eqn:ondapianaperturbata} \\
\displaystyle a(x,z) &=& a_{+}(z)e^{ikx} + a_{-}(z)e^{-ikx},
\label{eqn:perturbazioneoscillante}
\end{eqnarray}
with $|a(x,z)| \ll 1\;\forall x,z$. Using (\ref{eqn:ondapianaperturbata}) and (\ref{eqn:perturbazioneoscillante}) in (\ref{eqn:modifiedNLS}), we obtain the following system of differential equations
\begin{equation}\label{eqn:matriceperturb}
i\partial_{z}\left(
\begin{array}{c}
a_{+}\\
a_{-}\cc
\end{array}
\right) = \left(
\begin{array}{cc}
-R(k) -\gamma &  -\gamma   \\
\gamma           &  R^{\ast}(k) +\gamma
\end{array}
\right)\left(
\begin{array}{c}
a_{+}\\
a_{-}\cc
\end{array}
\right),
\end{equation}
with $R(k) = \frac{1}{\epsilon }\left( \sqrt{1-\epsilon k^2} -1 \right)$.
Eq.(\ref{eqn:matriceperturb}) can be solved on the basis of the eigenvectors;
correspondigly the solution can have an exponential evolution if the eigenvalues have a nonzero immaginary part. 
The eigenvalues $\lambda_{\pm}$ are
\begin{equation}
\lambda_{\pm} = -i R_{i} \pm |R_{r}|\sqrt{1-\frac{2\gamma}{|R_{r}|}},
\end{equation}
with $R_{r}(k)$ and $R_{i}(k)$ real and imaginary parts of $R(k)$. 

In a focusing medium ($\gamma >0$), wave-numbers such  $|R_{r}(k)| < 2\gamma$ exponentially growth w.r.t. $z$. 
The rate of the growth is given by the gain function $\lambda_{+}=i G$: 
\begin{equation}\label{eqn:gain1}
G(k) = \left\{
\begin{array}{ll}
\displaystyle \frac{1-\sqrt{1-\epsilon k^2}}{\epsilon}\sqrt{\frac{2\gamma\epsilon}{1-\sqrt{1-\epsilon k^2}}-1} &\quad |k|\le\frac{1}{\sqrt{\epsilon}}\vspace{2mm}\\
\displaystyle -\frac{1}{\epsilon}\sqrt{\epsilon k^2 - 1} + \frac{1}{\epsilon}\sqrt{2\gamma\epsilon -1} &\quad |k| >\frac{1}{\sqrt{\epsilon}}.
\end{array}
\right.
\end{equation}
The gain in Eq.(\ref{eqn:gain1}) has the expression found in the paraxial approximation,
in the limit $\epsilon\rightarrow 0$ ($k_{c}^{2} \equiv 4\gamma$):
\begin{figure}[t!]
\centering
\subfigure{\includegraphics[width = \columnwidth]{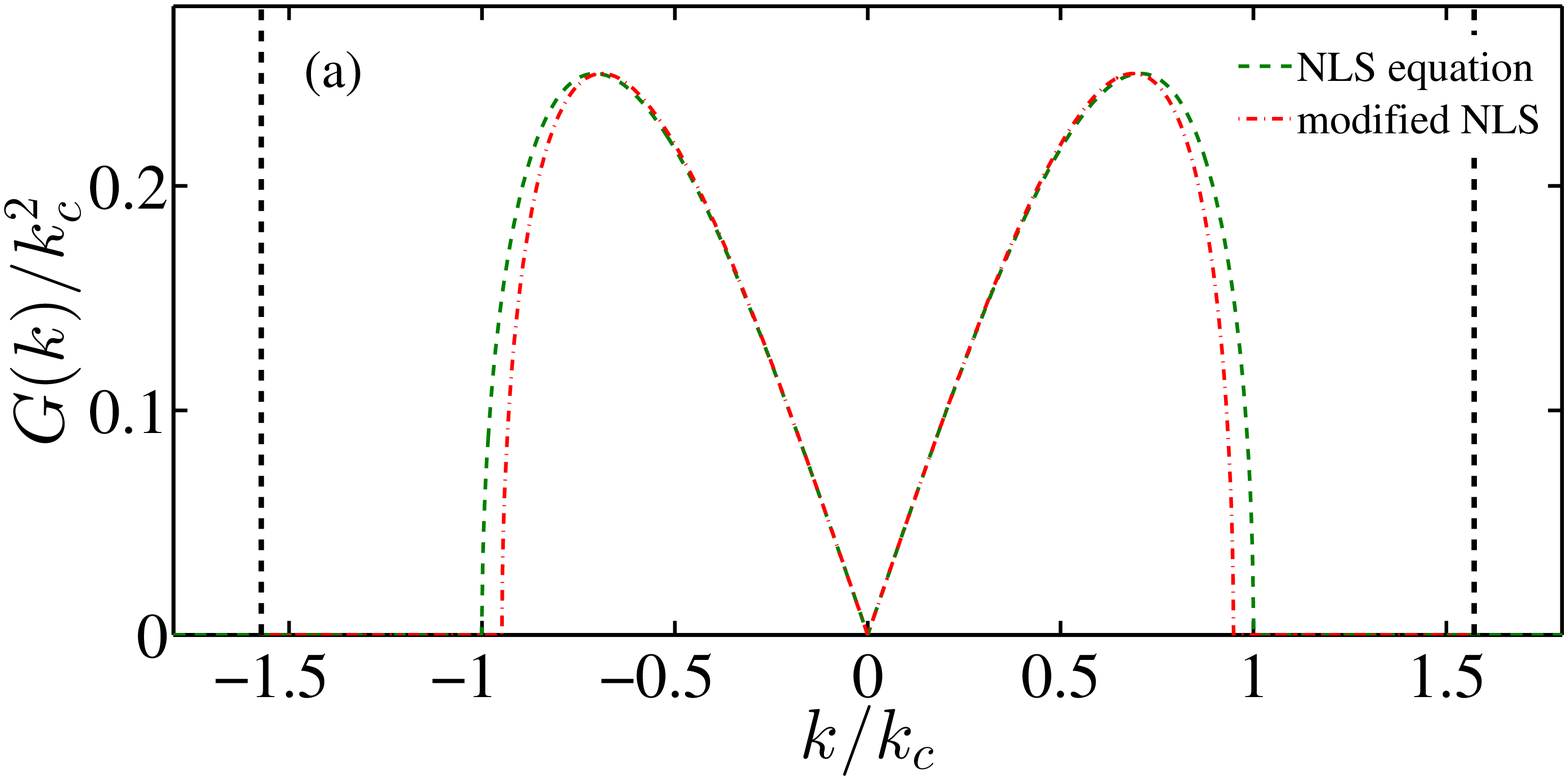}} \\
\subfigure{\includegraphics[width = \columnwidth]{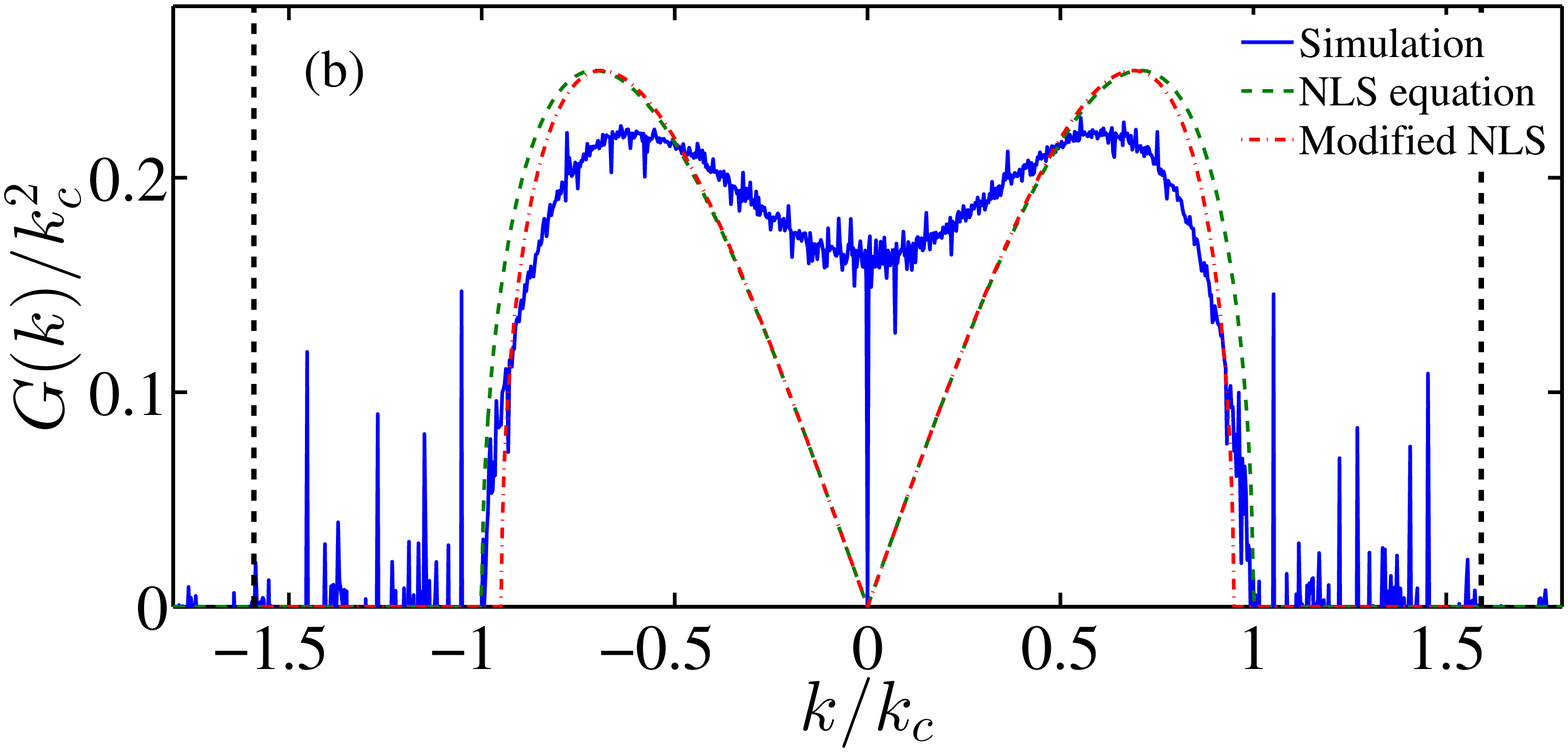}} \\
\subfigure{\includegraphics[width = \columnwidth]{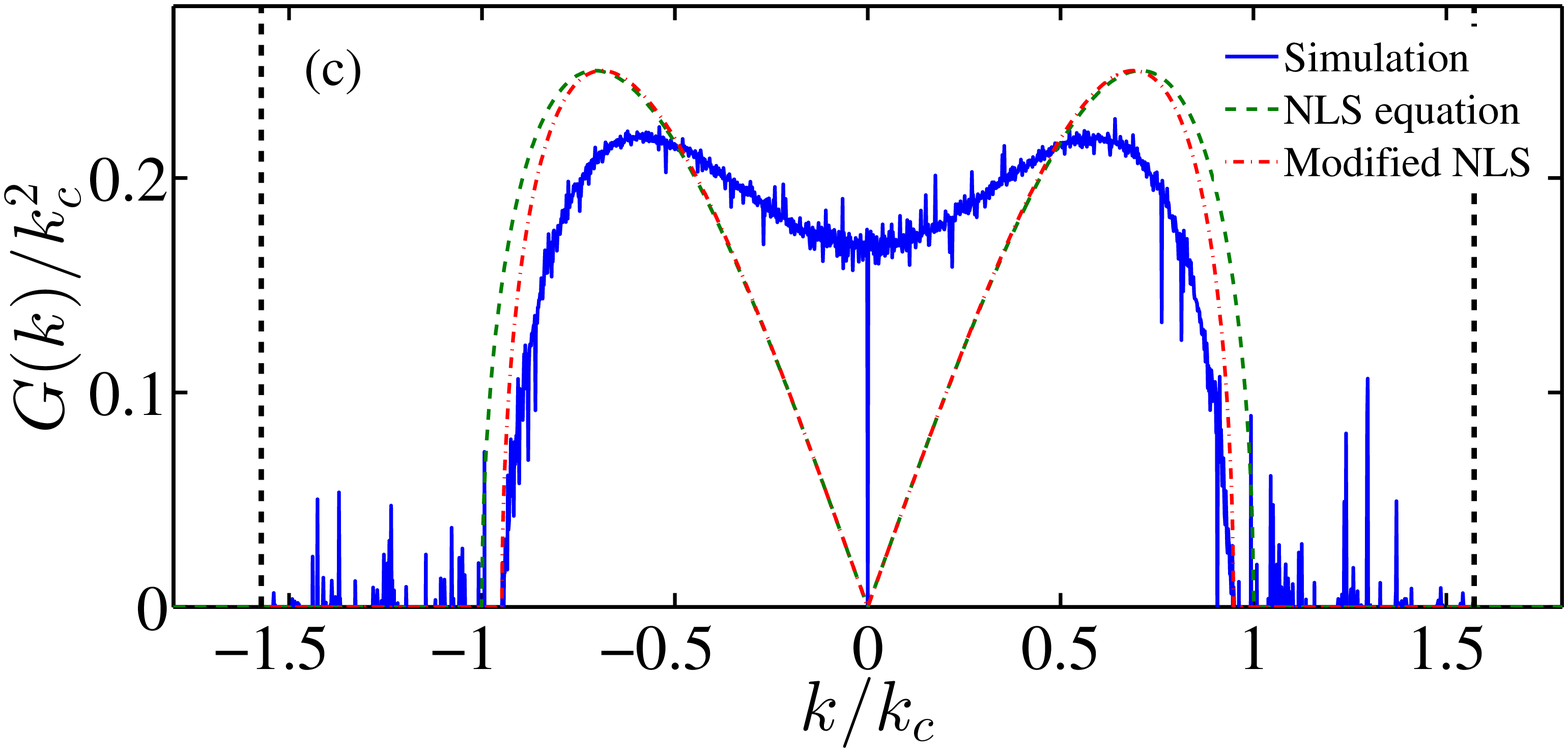}} 
\caption{(Color online) (a) Comparison of the theoretical gain functions from NLS and modified NLS.  
The choice of the scale for the axes in the figure is such that the gain function in the paraxial case is independent of the value of $\gamma$ while in the non-paraxial case it depends only on the product $\epsilon\gamma$. 
In panels (b) and (c) we show the comparison between the theoretical predictions and the numerical results for the gain function,
 finding a good agreement for the range of the amplified wave-numbers in the paraxial (b) and non-paraxial model (c). 
The simulations were performed by a split step Fourier method. The vertical dashed black lines separate the pro\-pagating Fourier components from the evanescent spectrum. The modified NLS equation takes correctly into account the non propagating nature of these components, indeed in (c) the noise beyond the black lines, present in (b), vanishes.
}\label{fig:gain nparax1} 
\end{figure}
\begin{equation}\label{eqn:gain}
G(k) = \frac{k^2}{2}\sqrt{\frac{k_{c}^{2}}{k^2}-1}.
\end{equation}

In Fig.\ref{fig:gain nparax1}a we compare the gain function for the pa\-raxial (\ref{eqn:gain}) and non-paraxial case (\ref{eqn:gain1}). 
The introduction of higher order diffraction terms leads to the reduction of the amplification bandwidth in the wave-number space: in the paraxial case, the instability region is given by $k<k_{c}$, while, in the non-paraxial case, 
is given by $k<k_{c}\sqrt{1-\epsilon\gamma}$. 

In addition, the maximally amplified wave-vector $k_{max} = k_{c}\sqrt{1-\epsilon\gamma/2}$ 
shifts towards lower wave-numbers with respect to the paraxial case $k_{max} = k_{c}/\sqrt{2}$, 
resulting in an increased wavelength of the modulation, i. e., the healing length $\Lambda = 2\pi / k_{max}$.
On the contrary, the maximum of the gain function is not affected by $\epsilon$. 

All these corrections are of the order $O(\epsilon\gamma)$, which means that they come from the coupling of the nonli\-near and non-paraxial effects. Therefore such corrections become more relevant when increasing the power (larger $\gamma$), or when decreasing the size of the beam (larger $\epsilon$). 

This is further confirmed by the fact that a different choice for the normalization parameters, 
i.e., expressing the transverse coordinates and the propagation distance in units, respectively, of the healing length $\Lambda$ and of the nonlinear length $L_{nl}$, leads to an UPPE only containing the single parameter $\epsilon\gamma$
\begin{equation}
i\partial_{z}\hat{\psi} + \frac{1}{\epsilon\gamma}\left(\sqrt{1-\frac{\epsilon\gamma}{2\pi ^2} k^2}-1\right) \hat{\psi} +{\cal P}[\psi]/\sqrt{1-\frac{\epsilon\gamma}{2\pi^2} k^2} = 0\text{.}
\end{equation} 

The physical meaning of the product $\epsilon\gamma$ is related to the ratio between the carrier wavelength $\lambda$ and the nonlinear length $L_{nl}$: indeed $\epsilon\gamma = 1/\beta_{0}L_{nl}$, where $\beta_{0} = 2\pi n_0/\lambda$. 
Since the nonlinear effects typically occur on a length scale much greater than that of the carrier wavelength of the beam, the corrections must be small $\epsilon\gamma \ll 1$. Otherwise, if $\epsilon\gamma \sim 1$, the envelope of the wave varies on the same length scale of the carrier wave and the use of envelope equations becomes meaning\-less. 
Therefore, we can retain as an upper bound the value $\epsilon\gamma\sim 0.1$. 

In real world experiments, the magnitude of the corrections is related to the nonlinear correction to the refractive index, indeed $\epsilon\gamma = \Delta n/n_{0}$ where  $\Delta n = n_{2}I_{0}$. 
Thus the value $\epsilon\gamma = 0.1$ can be obtained using highly nonlinear materials, such as liquid cristals \cite{Peccianti03, Peccianti2005} or thermal media \cite{Gentilini2013} and high intensity beams. 
As an example, one can consider a beam whose transverse width is $\sigma_{x} = 8.5\cdot 10^{-6} m$ with carrier wave number $\beta_{0} = 1.17\cdot 10^{7} m^{-1}$ and intensity peak $I_{0} = 1.5\cdot 10^{13} Wm^{-2}$ propagating in liquid cristals ($n_{2} \sim 10^{-14} m^2 W^{-1}$). The corresponding values of $\epsilon$ and $\gamma$ are respectively $10^{-4}$ and $10^3$ which lead to corrections of the order $\epsilon\gamma \sim 0.1$. 

In Fig.\ref{fig:gain nparax1}b,c we compare the gain function obtained from the simulation of the NLS equation (b) and the modified NLS equation (c) with the expected theoretical trend (a) for $\epsilon\gamma = 0.1$. The range of the amplified wave number is in quantitative agreement with the theo\-retical predictions, while the values of the gain function differs substantially in the region around $k\sim 0$. This is due to the numerical discretization which lead the initial spectrum to have a finite width instead of being a delta function. Thus the gain function can be thought as a convolution of the gain functions arising from the single Fourier components of the initial spectrum. In addition, as resulting from the theory, only the component of the initial field that is parallel to the direction of growth, identified by an eigenvector of the evolution matrix in equation (\ref{eqn:matriceperturb}), can be amplified. 

In order to calculate the gain function from the simulations we evaluate the amplitude of the Fourier spectrum $\hat{\psi}(z,k)$ at the initial point ($z = 0$) and after about two nonlinear length ($\gamma z \sim 2$), so the gain function is obtained from the numerical simulation through the simple relation $G(k) = \frac{1}{z}log(\hat{\psi}(k,z)/\hat{\psi}(k,0))$.

The evanescent waves are given by Fourier component with $k > 1/\sqrt{\epsilon}$, and decrease exponentially. This fact is correctly taken into account in the modified NLS equation as one can see from the theoretical results (\ref{eqn:gain1}) and simulations in Fig.\ref{fig:gain nparax1}c. This makes the model dissipative and it is valid until the corresponding losses of power remain negligible. 
We remark that the dissipation addressed here is not the known process due to material absorption, but it is an effective process due to the breaking of the uni-directional approximation, which does not produces a damping in the energy content of the beam, but a transfer to the backward component not included in the UPPE approach.
\subsection{\label{sec:breaking}Beyond modulation instability}
\begin{figure}[t!]
\centering
\subfigure{\includegraphics[width=0.49\columnwidth]{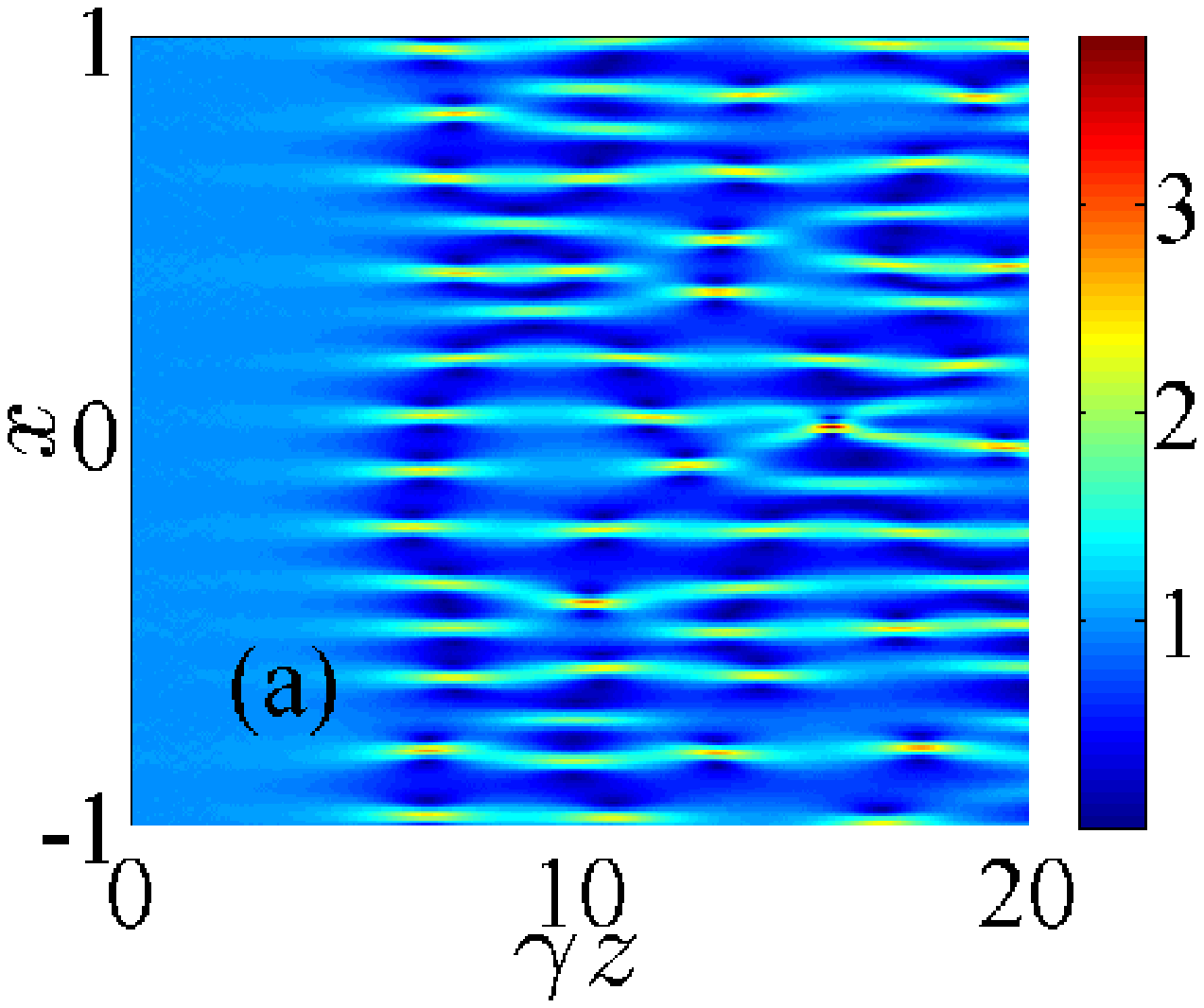}}  
\subfigure{\includegraphics[width=0.49\columnwidth]{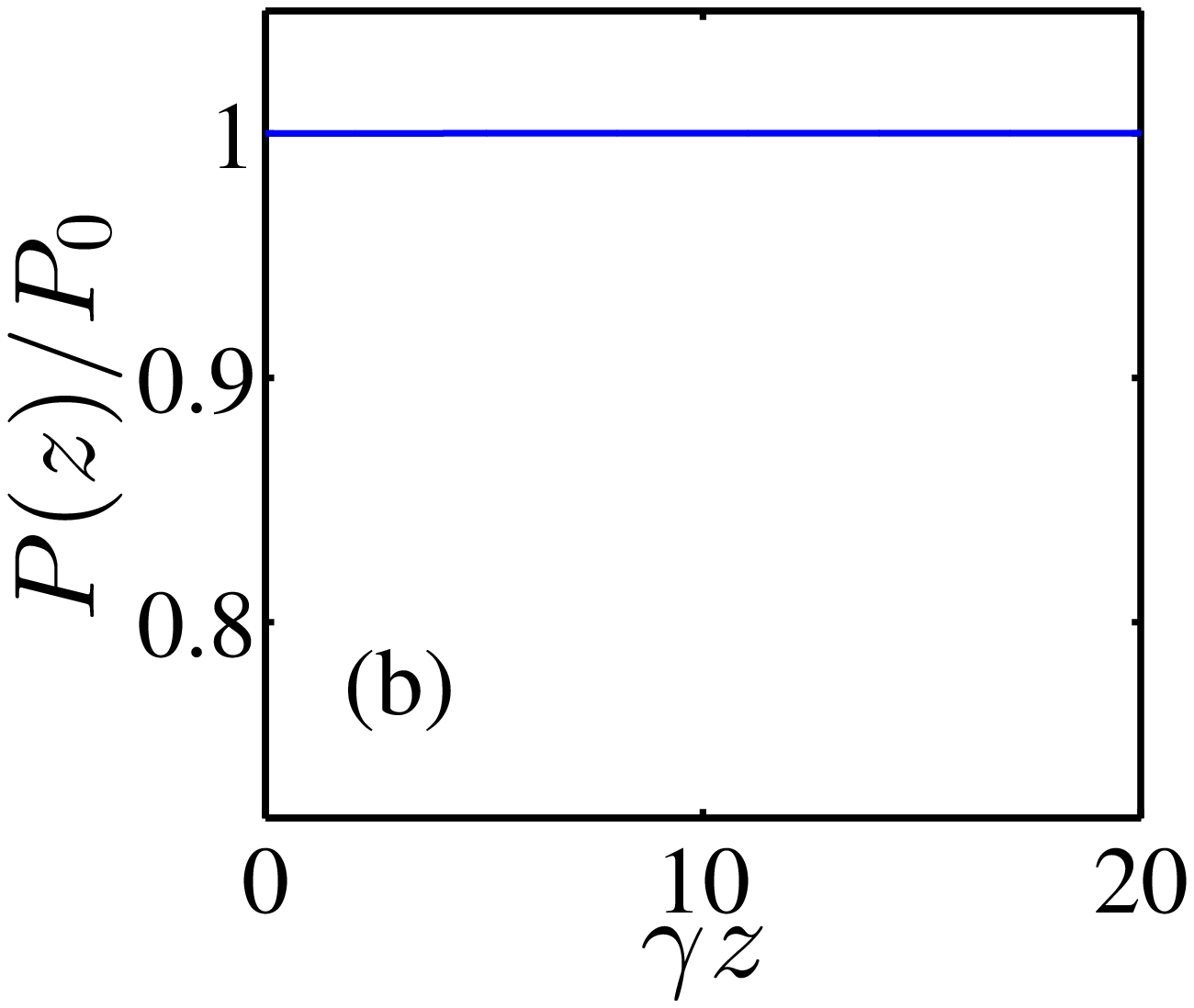}} \\
\subfigure{\includegraphics[width=0.49\columnwidth]{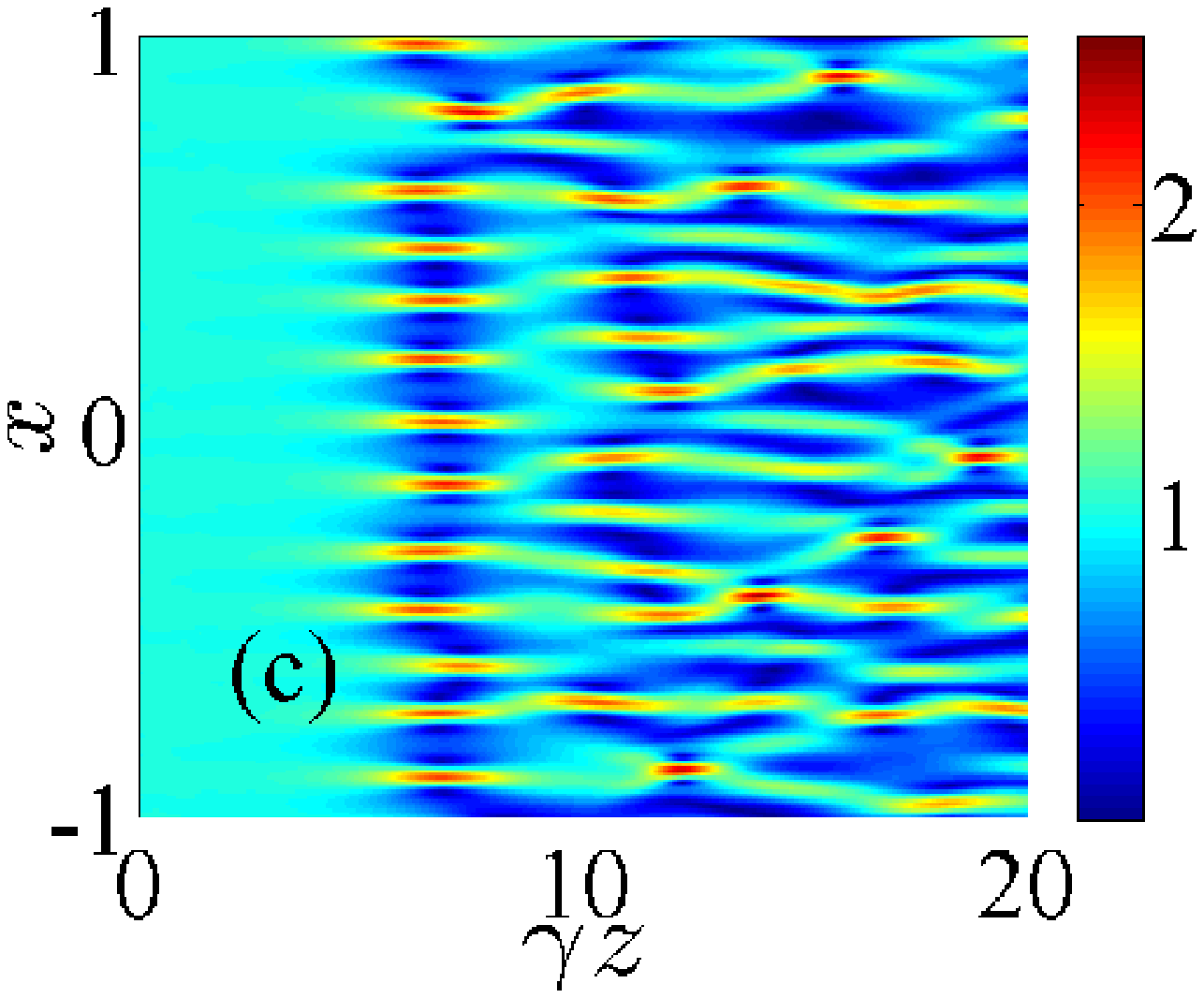}}
\subfigure{\includegraphics[width=0.49\columnwidth]{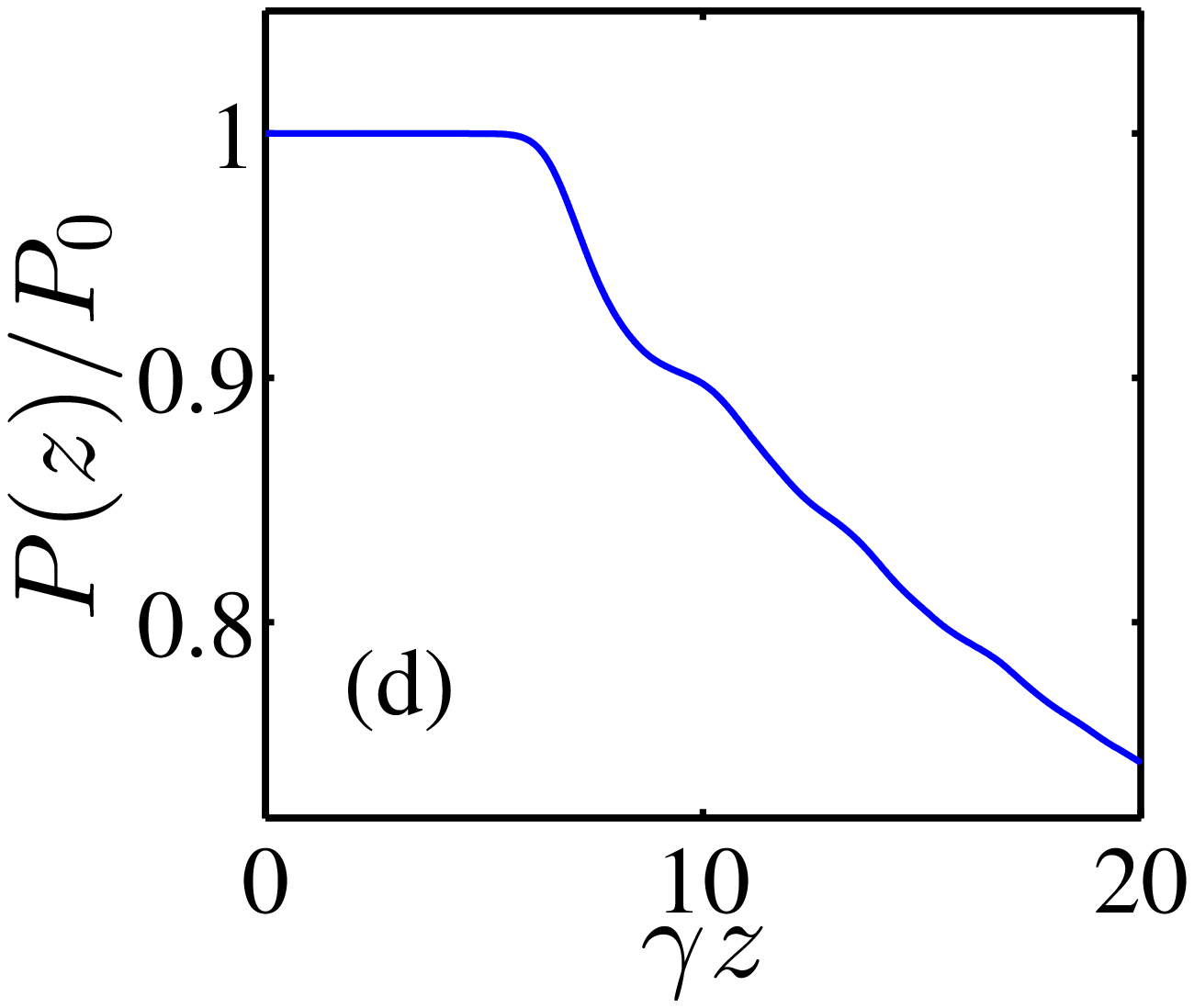}} 
\caption{(Color online) We simulate the filamentation process using the NLS equation (a,b) and the modified NLS equation (c,d). 
The initial condition is a constant of unitary amplitude perturbed by a small noise. The parameters used are $\epsilon = 10^{-4}$ and $\gamma = 10^3$. The filaments described by the modified NLS equation in panel (c) have wider size and less intensity in the region of focusing, with respect to the NLS equation in panel (a). This is due to the exponential decay of the evanescent components of the spectrum, which limits the narrowing of the filaments. The NLS equation preserves the power of the beam (b) while the modified NLS equation is a dissipative model (d) and hence it is valid until the relative loss of power can be considered negligible.
}\label{fig:break solitoni1} 
\end{figure}

MI can be observed after few nonlinear lengths. Going further along propagation, the initially constant amplitude of the beam, 
which has become modulated after two or three $L_{nl}$, breaks after six or seven $L_{nl}$ into localized structures, or ``filaments'', as in Fig.\ref{fig:break solitoni1}, which can be interpreted as bright solitons. 

The transverse distance between the filaments is of the order of magnitude of the healing length, which can be written in the paraxial case as $\Lambda = \sqrt{2\pi^2/\gamma}$. For exa\-mple, in Fig.\ref{fig:break solitoni1} we use $\gamma =10^3$ and the distance observed between the filaments is about $0.14$. 

These filaments focus and defocus periodically along the propagation. The main difference between the para\-xial and non-paraxial model regards the focusing which is more pronounced and localized in space for the former case in Fig.\ref{fig:break solitoni1}a; 
in the non-paraxial case the filaments di\-splay a larger minimum size and appear less focused, as shown in Fig.\ref{fig:break solitoni1}c. 
This is due to the exponential decay of the evanescent part of the spectrum that limits the narrowing of the filaments in the non-paraxial case. As we already pointed out, the exponential decay of the evane\-scent components makes the model dissipative, see Fig.\ref{fig:break solitoni1}d, while the NLS equation preserves the power of the beam along the propagation as in Fig.\ref{fig:break solitoni1}b. 
More specifically, when the filaments in Fig.\ref{fig:break solitoni1}c tend to focus, there is an enhancement of the amplitude of high 
wave-numbers in the spectrum, and the evanescent part decays exponentially causing a noticeable loss of power, as one can see from Fig.\ref{fig:break solitoni1}d. The evanescent components cannot propagate forward and hence they are reflected in the backward direction, which is not included in the UPPE and results as an effective loss mechanism. 

As long as the relative loss of energy is small, the mo\-dified NLS equation is a valid model for the amplitude of the forward propagating part of the electric field, other\-wise a more accurate model including also the backward propagating part of the field should be employed. 
Fig.\ref{fig:break solitoni1}d shows that the relative loss of power is less then $5\%$ within about seven nonlinear lengths ($\gamma z \sim 7$) and the modified NLS equation can be considered valid in the considered range. 

\subsection{\label{sec:Correction to the nonlinear term}Correction to the nonlinear term}
Here we consider the entire UPPE (\ref{eqn:UPPE_Psi}), that is we retain also the modifications to the nonlinear terms 
and investigate their effects. Since the model (\ref{eqn:UPPE_Psi}) exhibits two singularities for $k = \pm 1/\sqrt{\epsilon}$, we expand the nonlinear term to the first order in $\epsilon$ and consider only the relevant wave-numbers such that $|k| < 1/\sqrt{\epsilon}$;
indeed the assumptions $\epsilon \ll 1$ and $\epsilon\gamma \ll 1$ make the amplitude of the evanescent components very small 
and the range of the amplified wave numbers, which has a magnitude of the order of $k_{c}$, is largely contained 
in the interval $\sim 1/\sqrt{\epsilon}$. 

In the spatial domain the (1+1)UPPE can be written as
\begin{equation}\label{eqn:UPPE}
\left[i\partial_{z} + \frac{1}{\epsilon}\left(\sqrt{1  + \epsilon \partial^{2}_{x}} - 1\right)\right] {\psi} + \gamma|\psi|^2\psi - \frac{1}{2}\epsilon\gamma\partial_{x}^{2}(|\psi|^2\psi)= 0.
\end{equation}
\begin{figure}[t]
\centering
\subfigure{\includegraphics[width=\columnwidth]{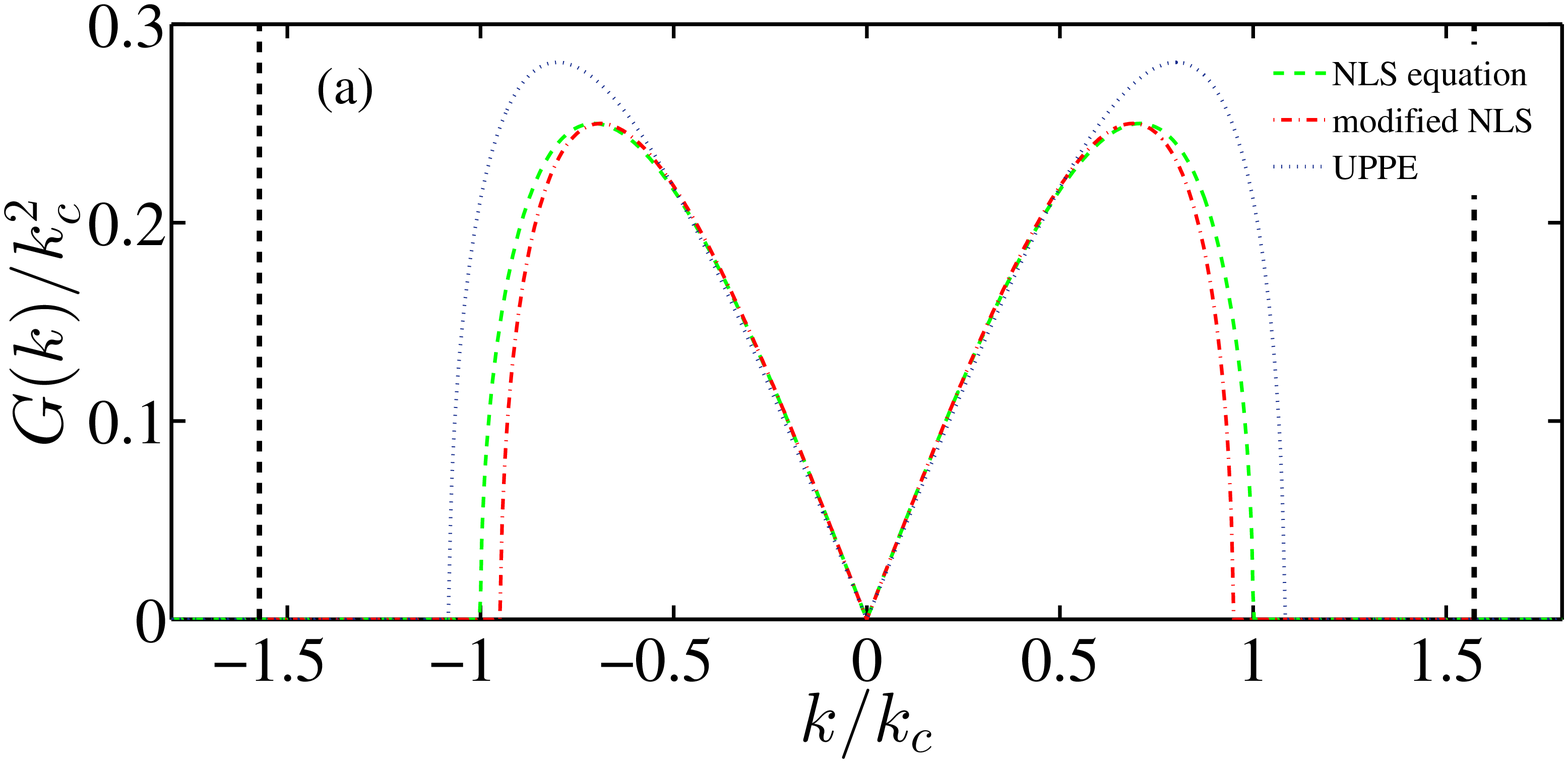}}  \\
\subfigure{\includegraphics[width=\columnwidth]{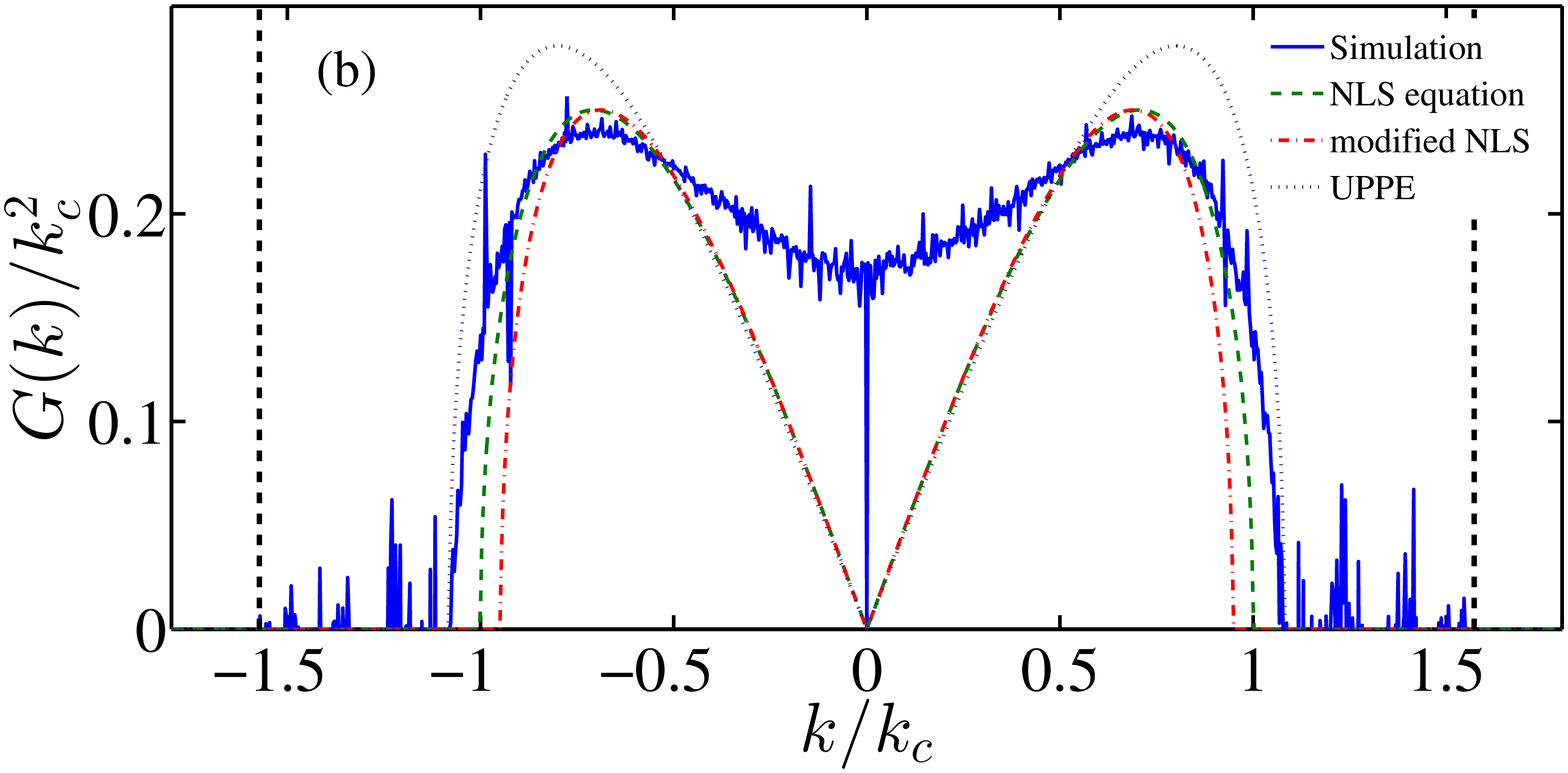}}   
\caption{(Color online) (a) Gain functions relative to the NLS, modified NLS and UPPE equations for $\epsilon\gamma =0.1$. (b) Comparison between the theoretical trends and the gain function obtained from the numerical simulation of the UPPE. }\label{fig:gain nparax2} 
\end{figure}
\begin{figure}[t!]
\centering
\subfigure{\includegraphics[width=0.49\columnwidth]{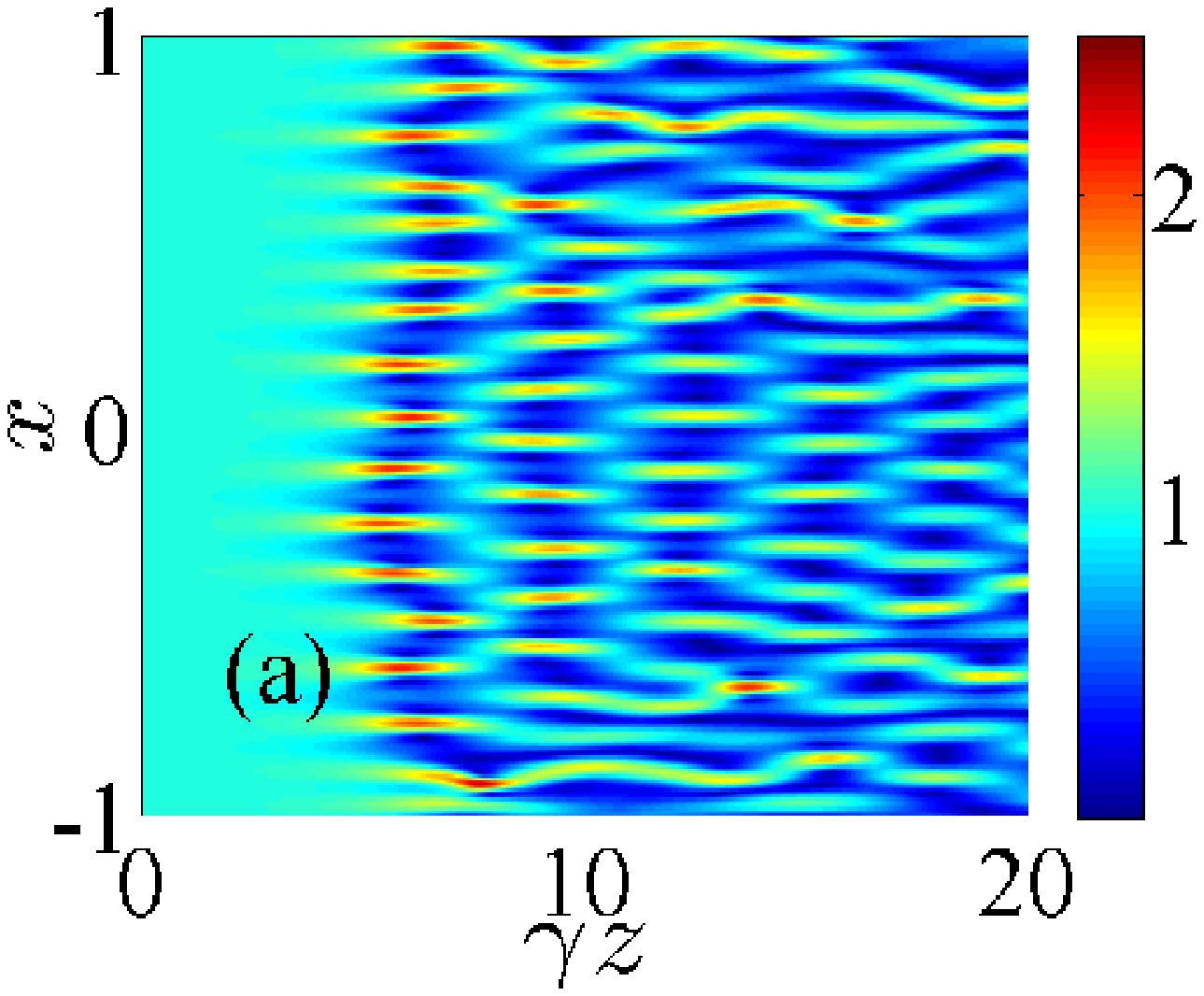}}  
\subfigure{\includegraphics[width=0.49\columnwidth]{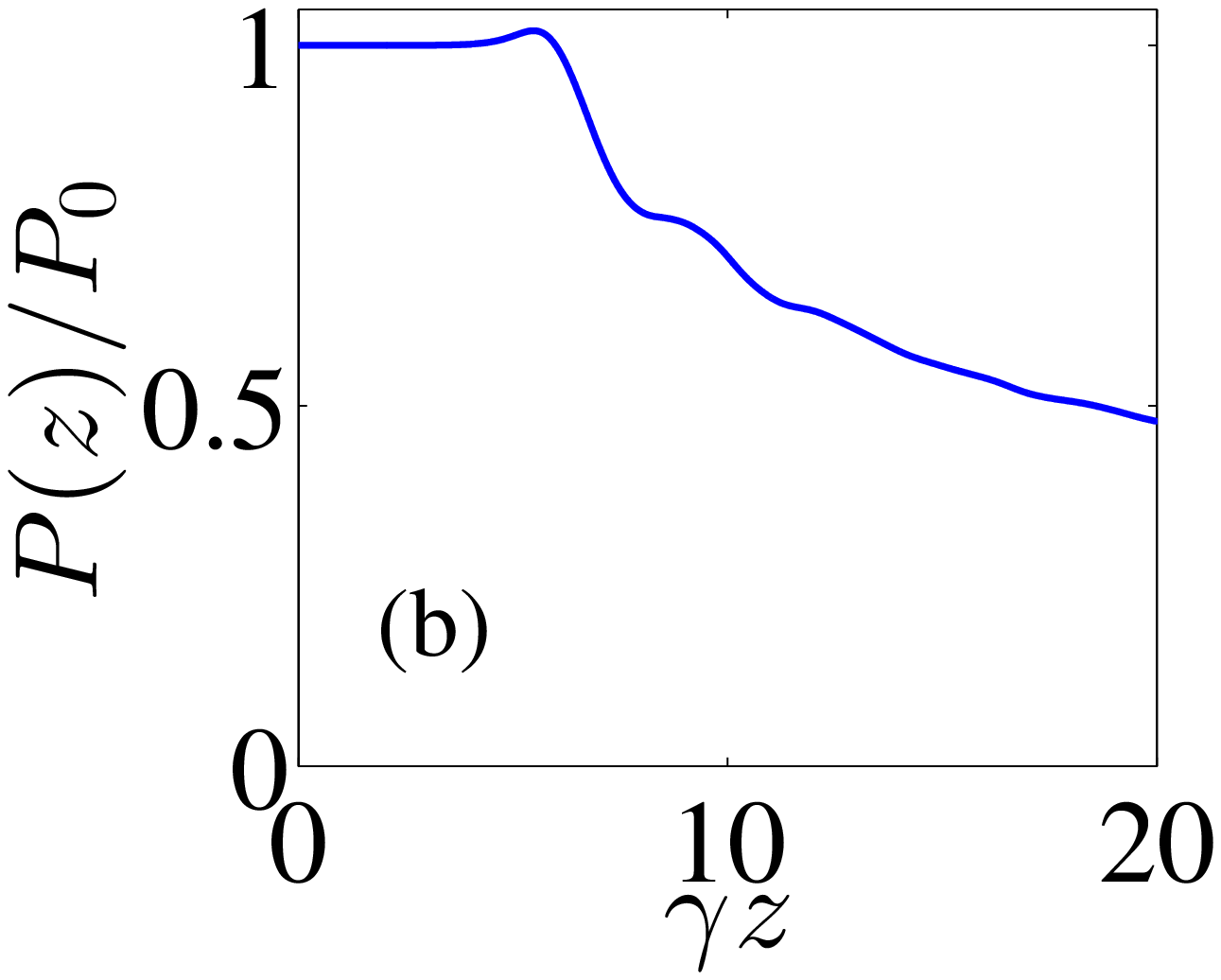}} \\
\caption{(Color online) (a) Filamentation process with the UPPE model. The filaments present the same structure as those in Fig.\ref{fig:break solitoni1}c . The introduction of small non-local interactions by the modification of the nonlinear term increases
the instability and this enhances the loss of power during propagation.}\label{fig:break solitoni2} 
\end{figure}
We note that the term $\frac{1}{2}\epsilon\gamma\partial_{x}^{2}(|\psi|^2\psi)$ introduces small corrections, which have a non-local character.
In order to calculate the gain function of the perturbation with this model we follow the same procedure above;
 we find the following expression for the gain function
\begin{eqnarray}\label{eqn:gain2}
\displaystyle G^2(k) &=&\displaystyle\left(\frac{1-\sqrt{1-\epsilon k^2}}{\epsilon}\right)^2\frac{2\epsilon\gamma}{1-\sqrt{1-\epsilon k^2}}(1+\epsilon k^2)+ \nonumber\\ 
& & \displaystyle -1 -\frac{(\epsilon\gamma)^2}{1-\sqrt{1-\epsilon k^2}}\left(\epsilon\frac{k^2 + \frac{3}{4}\epsilon k^4}{1-\sqrt{1-\epsilon k^2}}\right).
\end{eqnarray}
Eq.(\ref{eqn:gain2}) resembles (\ref{eqn:gain1}) for $k < 1/\sqrt{\epsilon}$. The difference is given by the presence of additional terms of the order $O(\epsilon^n \gamma)$ and $O(\epsilon^{n+1} \gamma^2)$ with $n\ge1$. These latter terms introduce higher order corrections that are negligible with respect to those considered in section \ref{sec:Corrections to the linear term} above; on the other hand, the former contain corrections of the order $O(\epsilon\gamma)$, which are to be taken into account.
These terms are responsible for the widening of the range of the amplified wave number, at variance with 
those in section \ref{sec:Corrections to the linear term}. In fact, performing the analysis to the first order in $O(\epsilon\gamma)$, we obtain for the amplification range $|k| < k_{c}\left(1+\frac{3}{2}\epsilon\gamma\right)$. Moreover, the wavenumber corresponding to the maximum of the gain function moves to higher values, so that the healing length gets smaller when increasing $\epsilon\gamma$. In addiction, 
as shown in Fig.\ref{fig:gain nparax2}, the values of the gain function grow and this enhances the instability process.

Fig.\ref{fig:break solitoni2}a shows the numerical simulation of the generation of filaments with the UPPE as a model. The fila\-ments have the same features of those relative to the modified NLS equation in Fig.\ref{fig:break solitoni1}c. The only difference is the loss of power along the propagation that is greater for the UPPE model because of the the greater instability induced by the nonlocal interactions, as shown in Fig.\ref{fig:break solitoni2}b. After about seven $\, L_{nl}$ the loss of power reaches the $20 \%$ of the initial power. In addition it is worth noticing that the correction to the nonlinear term introduces fluctuations in the energy that are of the order $O(\epsilon\gamma)$ and becomes more intense as the filaments get more focused. As a consequence one needs a more accurate model, which takes into account the back-propagating part of the field for the propagation of filaments after many nonlinear lengths.   

\begin{figure*}[t!]
\centering
\subfigure{\includegraphics[width=0.49\columnwidth]{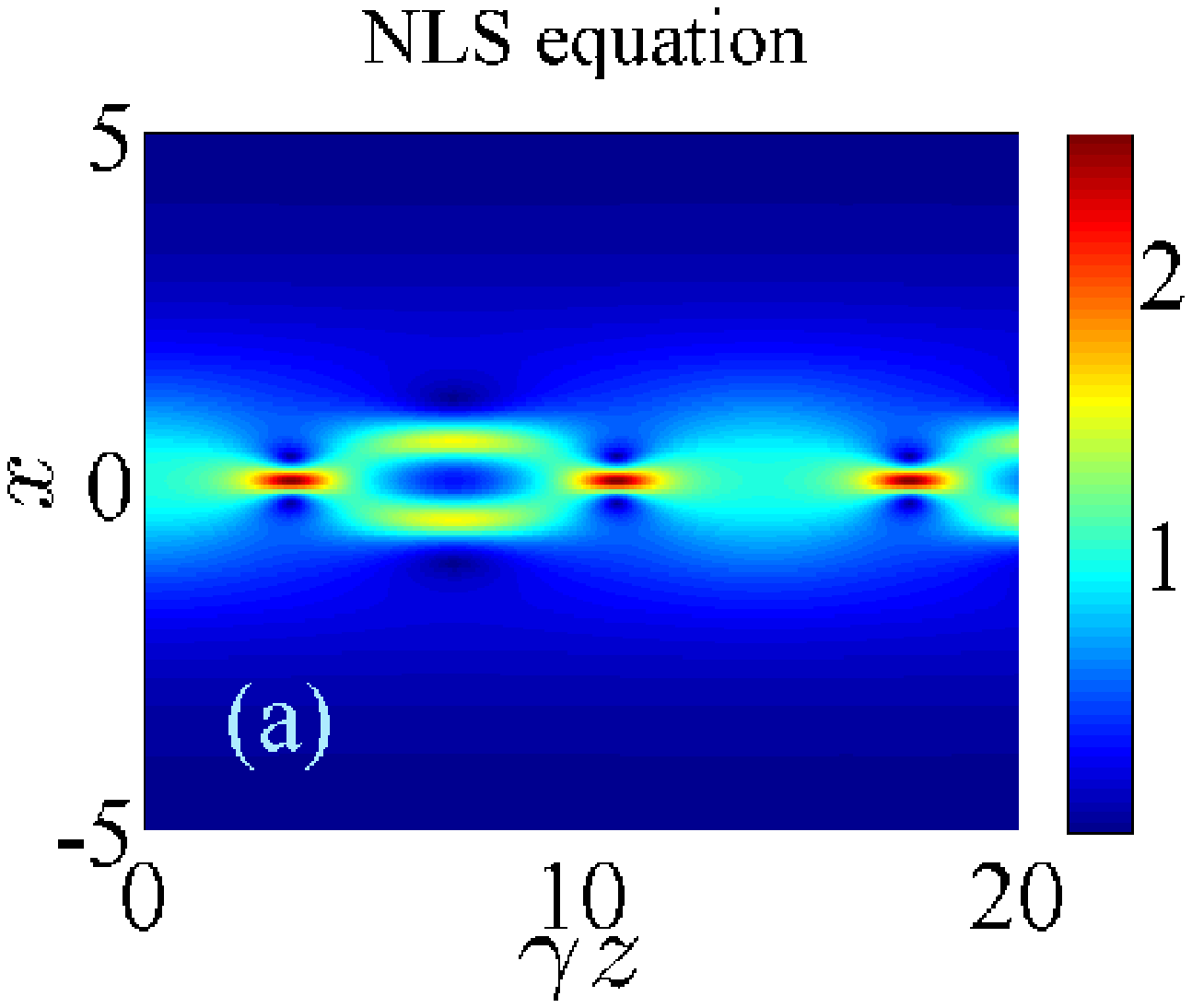}}  
\subfigure{\includegraphics[width=0.49\columnwidth]{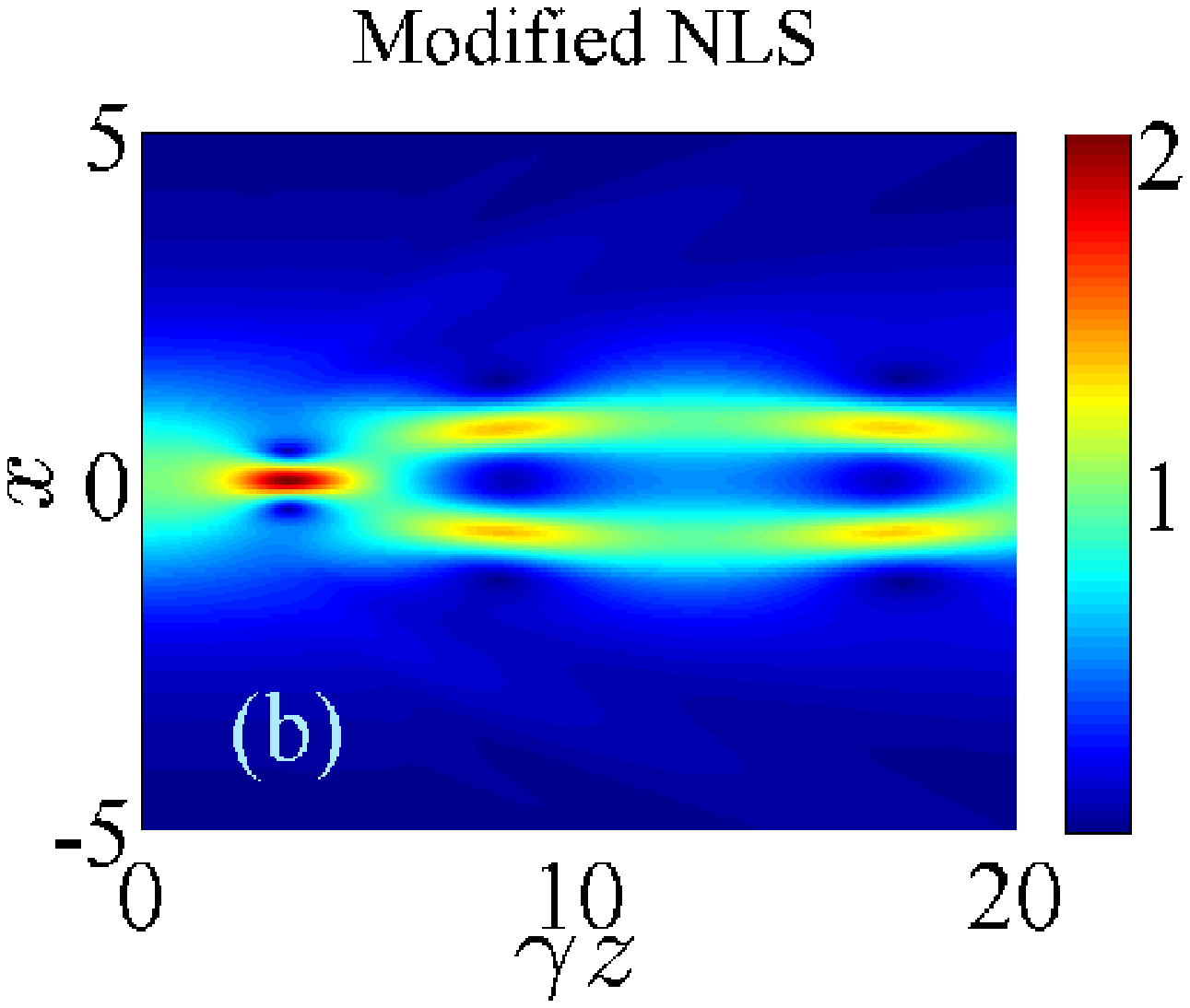}} 
\subfigure{\includegraphics[width=0.49\columnwidth]{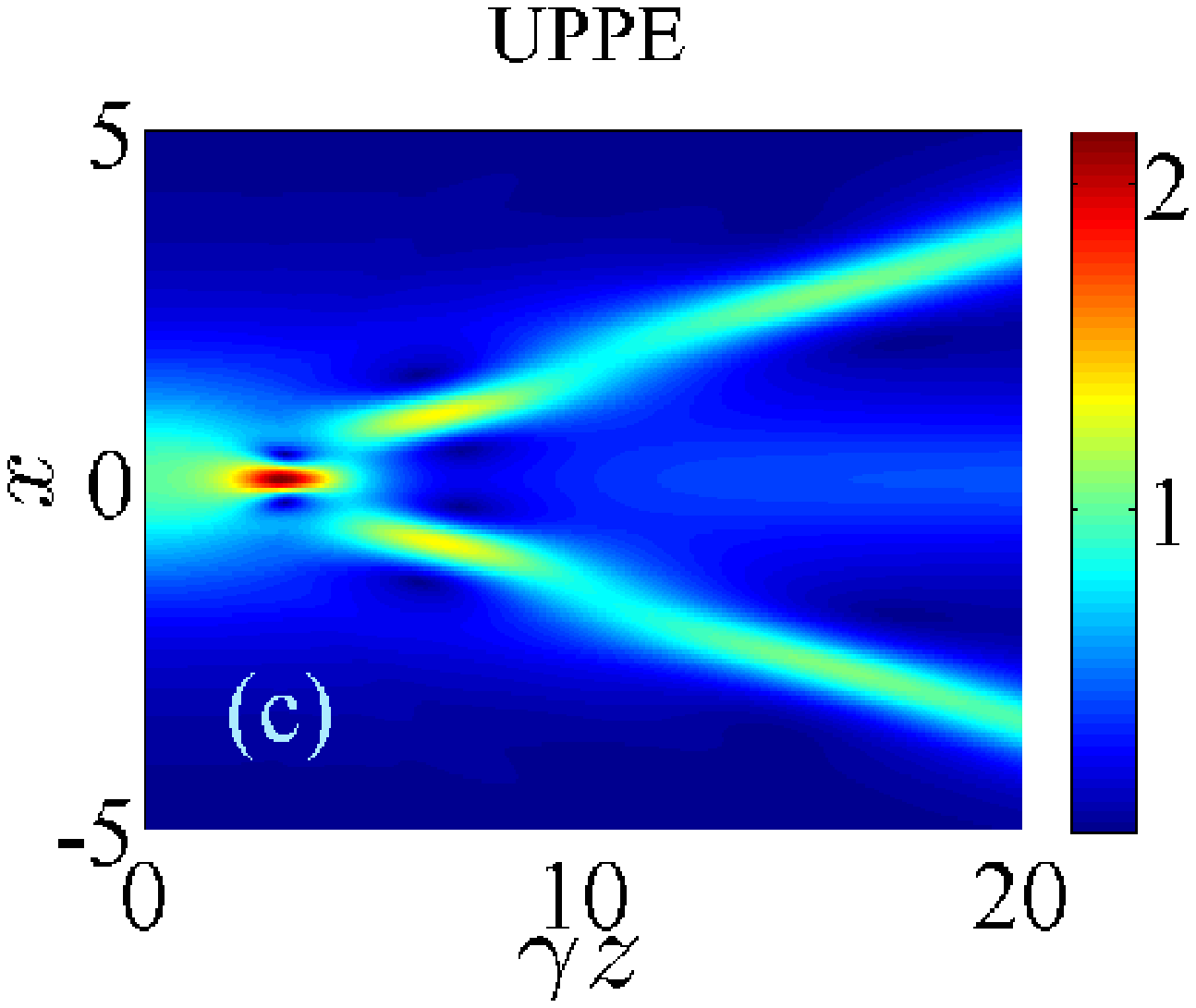}} \\
\caption{(a) The periodicity of a third order bright soliton ($\gamma =9$) is destroyed when including higher order corrections in the linear (b) and nonlinear term (c) of the NLS equation. The initial condition is $\psi(x,0) = sech(x)$ and the $\epsilon$ parameter is chosen as $\epsilon = 0.01$ such that $\epsilon\gamma \sim 0.1$. }\label{fig:solitons} 
\end{figure*}

Beyond the phenomenon of MI, the dissipative and non-local character of the corrections also affects the formation and the propagation of solitons. In fact the corrections breaks the Hamiltonian character of the NLS and hence also the periodicity of the soliton solutions. As an example we simulated the propagation of a third order bright soliton in Fig.\ref{fig:solitons}. We found that, during propagation the soliton, instead of describing a periodic pattern as in the paraxial model, Fig.\ref{fig:solitons}a, breaks in two distinct structures, Fig.\ref{fig:solitons}c. Again we find that the loss of power is noticeable, so a more accurate description of the propagation of a non-paraxial third order soliton must involve the back propagating part of the field.

\section{\label{sec:Conclusions}Conclusions}
We have considered the process of the modulation instability of a coherent laser beam propagating in a nonlinear 
Kerr medium beyond the ordinary NLS equation,  and employing the UPPE \cite{Moloney2004}. 
We considered the va\-rious corrections to the NLS equation starting from the ones to the linear terms and then including modifications
 to the nonlinear part. The additive linear terms account for the effects of higher order diffraction 
and for the evane\-scent Fourier components. We analyzed the way such corrections influence the propagation of a laser beam in a linear medium. 
It turns out that the beam diffracts more rapidly than in the paraxial case, as it can be described by closed form expressions, 
with results in perfect agreement with the numerical simulation. 

We then considered the modulation instability of a plane wave solution in a nonlinear Kerr medium and compare 
the results of the paraxial model with the ones from the modified NLS equation with non-paraxial corrections to the linear term. We found that the differences in the MI gain function are of the order of $\epsilon\gamma$,  that is, the corrections to the paraxial case raise from the coupling between the non-paraxial and nonlinear effects. Then introducing also the non-paraxial correction to nonlinear term, 
which has non-local features, the gain function is further modified and in particular it grows by an amount of the order of $\epsilon\gamma$. 

The product $\epsilon\gamma$ is related to the ratio between the carrier wavelength of the electric field and the nonlinear length. This product can be simply written as the ratio between the nonlinear modification of the refractive index and the linear refractive index, i.e. $\epsilon\gamma = \Delta n/ n_{0} = n_{2}I_{0}/n_{0}$, and so $\epsilon\gamma \ll1$. Thus the corrections to the NLS behavior are evident only in highly nonlinear media such liquid crystals \cite{Peccianti03, Peccianti2005} or thermal media \cite{Gentilini2013}, and with high intensity beams such that $\epsilon\gamma\sim 0.1$. 

In addition, the modification to the linear term is such that the model is able to treat correctly the evanescent Fourier components of the spectrum, which decay exponentially introducing dissipation in the model due to the fact that, as the spectrum of the wave broadens, a certain amount of energy is transferred to the evanescent waves, which excite backward propagating waves. As long as the relative loss of power is small the UPPE is a valid model that provides an accurate description for the propagation of a coherent laser beam, otherwise one has to include in the theory the backward propagating part of the electric field and the interaction between the forward and the backward part.

In the regime beyond the linear stage of MI, we investigated the generation of filaments and the effect introduced by the considered corrections. We found that the focusing of filaments is less pronounced in the non-paraxial case. This is due to the cut off of the Fourier components in the Fourier space that limits the focu\-sing in the real space. The great loss of power after about seven nonlinear lengths indicates that a more accurate model which involves the backward propagating part of the field is indeed needed to describe better the physics beyond
 several nonlinear lengths. In addition non-paraxiality breaks the periodicity of the soliton solutions because of the dissipative and non-local character of the corrections.


\providecommand{\noopsort}[1]{}\providecommand{\singleletter}[1]{#1}%

\end{document}